# Galilean invariant retarded electric interaction of moving charges

**Balázs Vető**,

Retired associate professor of Eötvös University, Budapest, Hungary

*e-mail: veto@metal.elte.hu*

**Abstract**

A Galilean invariant description of the retarded electric interaction of two moving charges is carried out, considering the EM action to propagate at the speed of light. Through the altering of the Liénard-Wiechert potential (LWP) a new retarded, electric potential called retarded Coulomb potential (RCP) is ordered to the moving charges. RCP depends on the velocities both of the interacting charges. The force law of the electric interaction of moving charges is determined by means of the second order approximation of RCP. The force law obtained is the Galilean invariant Weber's force law, surprisingly. The rediscovery of the Weber's force law from a retarded electric potential confirms its significance and proves it to be an approximated law, based on retardation. From the fact that the purely electric RCP provides not only the electric, but even the magnetic forces, a substantial ascertainment follows: In Galilean invariant description, magnetic phenomena are the manifestation of the retarded electric interaction exclusively.

## 1. Introduction

### 1.1 The principle of objectivity in physical laws, Galilean and Lorentz invariance

The principle of objectivity is a basic rule of formulating laws of physical phenomena. Objectivity requires the physical laws not to depend on the state of motion of the observer. This requirement is confined to observers fixed in inertial frames of reference (IFR) in Newtonian mechanics. When a physical law takes same form in all IFR in Euclidean geometry it fulfills the principle of objectivity and it called a Galilean invariant law. At the early stage of the electromagnetism Weber [1] gave a Galilean invariant description of the electromagnetic (EM) interaction which implies the relative position and relative velocity and acceleration of the interacting objects only. Constructing his law Weber considered an action-at-a-distance interaction between moving charges, as it was usual in Newtonian physics.

Maxwell's field theory [2] contradicts with the Galilean covariance from its outset. The magnetic field of a moving charge introduced by Maxwell's implies the velocity of its source charge. This velocity depends on the choice if the IFR. This way, the covariant description of Maxwell's field theory cannot be based on Galilean invariance. The covariance problem of Maxwell's field theory was solved by special relativity (SR), which is based on Minkowski geometry and Lorentz transformation. Maxwell's field theory satisfies the requirement of objectivity by Lorentz invariant formulation of laws.

Maxwell's field theory has one more feature. In contrast with Newton and Weber, Maxwell considers the EM action to propagate at the speed of light instead of the action-at-a-distance. In this approach, the retardation must also be taken into account when formulating the force law of two moving charges.

### 1.2 Poincaré's principle

In today physics the only way to describe the EM interaction of moving charges is SR and Minkowski geometry. In contrast with this obligate method Poincaré [3] stated that the geometry ordered to the physical space is not a property of the space, but a product of the human mind and it is arbitrarily adapted to the space. Hereby, different geometries can be ordered to the physical space. Physical laws can be formulated with identical meaning, in any geometry (with curtain properties), while the mathematical form of the laws and the definition of the affected physical quantities depends on the geometry. This ascertainment is Poincaré's principle. It suggests that a covariant description of the retarded EM interaction might be formulated not only in Minkowski, but even in Euclidean geometry – by satisfying Galilean instead Lorentz invariance.

*1.3 The programme of this study*

In this paper a Galilean invariant force law of the electric interaction of two moving charges is derived. The force law takes the propagation speed of EM action to be the speed of light and is required to range with the experience. To perform the derivation, a retarded Coulomb potential (RCP) of two moving charges is needed. Unfortunately, LWP is valid only in the rest frame of the charge which one perceives the potential. RCP is a retarded, pure electric potential which extends its validity for two moving charges. RCP will be obtained by the revision of the LWP and by taking some refinements into account. The sought after Galilean invariant force law comes from the second order approximation of the RCP.

## 2.  The electromagnetic interaction of moving charges

*2.1 Action-at-a-distance EM interaction and Weber's force law*

The first law of force between electrified bodies at rest originates from Coulomb [4], [5] in 1788. The law is wearing Coulomb's name. Ampére [6] formulated a quantitative law of magnetic force between elements of two current carrying wires in 1826, called Ampére's law. By unifying the Coulomb's and the modified Ampére's laws into one equation Weber implemented [1] the general law of EM forces in 1846:

$$\mathbf{F}_{1W} = \frac{kq_1q_2}{r^3}\mathbf{r}\left[1 + \frac{v^2}{c^2} - \frac{3(\mathbf{v}\mathbf{r})^2}{2r^2c^2} + \frac{\mathbf{a}\mathbf{r}}{c^2}\right] = -\mathbf{F}_{2W}. \tag{2.1}$$

The $\mathbf{F}_{1W}$ stands for the electromagnetic force exerted by $q_2$ on $q_1$. The vector $\mathbf{r} = \mathbf{r}_1 - \mathbf{r}_2$ points from $q_2$ to $q_1$, $\mathbf{r}_1$ and $\mathbf{r}_2$ are the time dependent position vectors of $q_1$ and $q_2$, $k = \frac{1}{4\pi\varepsilon_0}$ is an SI constant. Bold letters denote vectors and slanted ones their absolute value; $r = |\mathbf{r}|$. The vectors $\mathbf{v}$ and $\mathbf{a}$ are the first and second order derivatives of $\mathbf{r}$ by time, i.e. they stand for relative velocity and acceleration of $q_1$ and $q_2$. Hereafter this notation will be used generally.

Weber's force law describes the EM interaction of moving point charges and current carrying conductors in agreement with experience. In conflict with Maxwell's electrodynamics it predicts a magnetic force between a charge at rest and a current carrying wire. The existence of this force is not confirmed neither refuted experimentally, Assis and Hernandes [7]. Weber's law is Galilean invariant. It predicts identical forces in all inertial frames of reference and satisfies the law of action and reaction. Weber confirmed his law by performing precise measurements [1]. The coupling constant $\frac{1}{c^2}$ of electric and magnetic terms in (2.1) was determined by Weber and Kohlrausch in 1856 experimentally as Assis and Torres [8] refer. They found $c^2$ to be the square of speed of light. Weber also determined the interaction energy from which his force law (2.1) originates,

$$U = \frac{kq_1q_2}{r}\left[1 - \frac{(\mathbf{v}\mathbf{r})^2}{2r^2c^2}\right]. \tag{2.2}$$

In 1871 Weber proves [9] that his force law (2.1) satisfies the principle of conservation of energy. He took notice of a small increase ($\varepsilon$) of the mass of electrically charged particles in electric interaction, but did not manifest its rate. Today we know this rate; $\varepsilon = kq_1q_2/(rc^2)$, it is the mass of the Coulomb interaction energy. Weber's law was formulated in the framework of Newtonian physics. Using Newtonian approach Weber considered an action-at-a-distance character of EM interaction.

With the successful Hertz's experiments [10] the existence of EM waves became evident and the action-at-a-distance untenable. Therefore Weber's law disappeared from the textbooks and from the scope of mean stream physical research for the end of the 1920-s.

*2.2 Retarded EM interaction, the Liénard-Wiechert potential (LWP) – potential of a moving charge*

A basic element of Maxwell's field theory [2] is the revelation of the finite propagation speed of the EM action. The finite propagation speed causes retardation in EM interactions and this delay modifies the EM potentials, fields and forces. Maxwell did not formulate the effect of retardation mathematically. It was done later, after Maxwell's death by Liénard [11] and Wiechert [12]. This retarded potential is thus referred as Liénard-Wiechert potential. LWP determines the retarded potential around a moving point charge in an observation point at rest.

LWP was calculated in Euclidean geometry. Let $q_2$ be the point charge which generates the potential in the observation point $P_1$ and is moving along the trajectory $\mathbf{r}_2(t)$. The position vector $\mathbf{r}_2(t)$ points from point $O$ to $q_2$. The constant position vector $\mathbf{r}_1$ points from point $O$ to $P_1$. See figure 1. The position vectors $\mathbf{r}(t) = \mathbf{r}_1 - \mathbf{r}_2(t)$ and $\mathbf{r}(t_r) = \mathbf{r}_1 - \mathbf{r}_2(t_r)$ point from $q_2$ to $P_1$ at the instants $t$ and $t_r$, respectively. The action starts from $q_2$ at the moment $t_r$, and get to $P_1$ just at $t$.

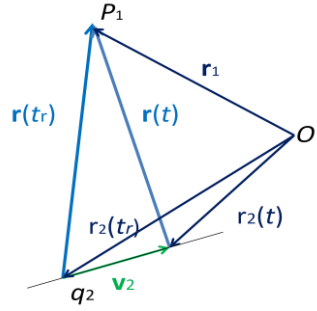

**Figure 1.** *Retarded LW potentials generated by the moving $q_2$ charge in the arbitrary observation point $P_1$.*

The EM action needs time to reach from the source to the observation point. Using that the EM action propagates at the speed of light and considering that the motion of $q_2$ does not influence the light's propagation speed. The length of the retardation time is,

$$t - t_r = \frac{r(t_r)}{c}. \tag{2.3}$$

The LWP-s of the moving charge $q_2$ in the observation point $P_1$ at the moment $t$ are:

$$\varphi_{2LW}(\mathbf{r}_1, t) = \frac{kq_2}{r(t_r) - \frac{\mathbf{v}_2(t_r)\mathbf{r}(t_r)}{c}}, \qquad \mathbf{A}_{2LW}(\mathbf{r}_1, t) = \varphi_{2LW}(\mathbf{r}_1, t)\frac{\mathbf{v}_2}{c^2} = \frac{kq_2}{r(t_r) - \frac{\mathbf{v}_2(t_r)\mathbf{r}(t_r)}{c}}\frac{\mathbf{v}_2}{c^2}. \tag{2.4}$$

In the denominator of LWP (2.4) there is a correction term added to the distance of $r(t_r)$ which implies the velocity of the source charge: $-\mathbf{v}_2(t_r)\mathbf{r}(t_r)/c$. This way LWP is not Galilean invariant. The correction term can be interpreted as the length of the radial displacement of the source charge, $q_2$ in the direction of $\mathbf{r}(t_r)$ at a velocity of $\mathbf{v}_2(t_r)$ within the retardation time: $t - t_r$.

*2.3 The meaning and the effect of the arbitrary restriction used in the derivation of LWP*

Maxwell's field theory supposes that a moving charge possesses a potential depends on its position and velocity. In order to have LWP only depend on the velocity of the source charge, the observation point $P_1$ required be at rest. The condition, $\mathbf{v}_1 \equiv 0$, keeps the retarded potential in the frame of Maxwell's field theory. The constancy of the vector $\mathbf{r}_1$ (figure 1.) has an important effect in the determining of the correction term. In the derivation of LWP [13] the correction term originates from the partial derivative;

$$|\mathbf{r}_1 - \mathbf{r}_2(t_r)|\frac{\partial}{\partial t'}\left[t - \frac{1}{c}|\mathbf{r}_1 - \mathbf{r}_2(t')|\right]\Big|_{t'=t_r} = -\frac{\mathbf{r}_1 - \mathbf{r}_2(t_r)}{c}[\mathbf{v}_2(t_r)]. \tag{2.5}$$

Here the vectors $\mathbf{r}$ and $\mathbf{r}_s(t')$ used in [13] are replaced by the positions of the observer $\mathbf{r}_1$ and the source $\mathbf{r}_2(t')$, fittingly to our notation. When the observation point is allowed to move then $\mathbf{r}_1$ becomes a function of $\mathbf{r}_1(t')$ too. Using the notation of relative positions and relative velocities of the observing point and of the source charge, $\mathbf{r}(t_r) = \mathbf{r}_1(t_r) - \mathbf{r}_2(t_r)$ and $\mathbf{v}(t_r) = \mathbf{v}_1(t_r) - \mathbf{v}_2(t_r)$ then the correction term (2.5) takes the corrected form of;

$$|\mathbf{r}_1(t_r) - \mathbf{r}_2(t_r)| \frac{\partial}{\partial t'}\left[t - \frac{1}{c}|\mathbf{r}_1(t') - \mathbf{r}_2(t')|\right]\Big|_{t'=t_r} = \frac{\mathbf{r}_1(t_r) - \mathbf{r}_2(t_r)}{c}[\mathbf{v}_1(t_r) - \mathbf{v}_2(t_r)] = \frac{\mathbf{r}(t_r)\mathbf{v}(t_r)}{c}. \quad (2.6)$$

An additional effect of the motion of the observation point that the retarded time $t_r$ (2.3) changes to $t'_r$:

$$t - t'_r = \frac{s_2(t'_r)}{c}, \quad \text{where} \quad \mathbf{s}_2(t'_r) = \mathbf{r}_1(t) - \mathbf{r}_2(t'_r). \quad (2.7)$$

(See figure 2.) Whereupon the correction term (2.6) takes the next from of;

$$\frac{\mathbf{v}(t'_r)\mathbf{r}(t'_r)}{c} = \frac{t - t'_r}{r(t'_r)}\mathbf{v}(t'_r)\mathbf{r}(t'_r). \quad (2.8)$$

In the case of a moving observation point the relative velocities of the source and the observer appears in the correction term of the retarded potential. A potential which involves the observers velocity is outside of Maxwell's field theory. That is the reason of the restriction of $\mathbf{v}_1 \equiv 0$. Because of this restriction LWP is valid only in the rest frame of the observation point and therefore cannot be applied for interaction of two moving charges.

## 3   A Galilean invariant approach to the retarded electric potential of moving charges

### 3.1 Making LW valid for two moving charges

As in the above section it was established LWP has an arbitrary restriction; the observation point of the LWP must be at rest. In section 2.3 was also shown, that in the case of a moving observation point its velocity appears in the correction term (2.8) of the retarded potential. The length of the retardation time changes as well.

To get rid of the arbitrary restriction of $\mathbf{v}_1 \equiv 0$, LWP will be modified to imply the relative position and velocity of the interacting charges. It will be named corrected LWP (CLWP). The investigation of the retarded electric interaction is intended hence the only scalar term of LWP will be used.

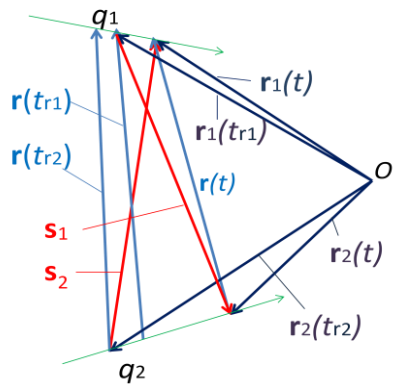

***Figure 2.*** *Geometric representation of two moving charges.*

The CLWP potentials $\varphi_{2CL}(\mathbf{r}_1, t)$ and $\varphi_{1CL}(\mathbf{r}_2, t)$ which are generated and perceived by $q_1$ and $q_2$ at each-other location conversely, will be formulated. The action perceived by the charges $q_1$ and $q_2$ at the moment $t$, starts from their sources at different $t_{r1}$ and $t_{r2}$ instants. The electric action traces different $s_1$ and $s_2$ distances coming along from the sources $q_1$ and $q_2$ to the perceivers $q_2$ and $q_1$, see figure 2. Due to the symmetry, the formulation of the potentials $\varphi_{2CL}(\mathbf{r}_1, t)$ and $\varphi_{1CL}(\mathbf{r}_2, t)$ follows the same steps, so it is enough to derive only one of them. Let us here choose the calculation of $\varphi_{2CL}(\mathbf{r}_1, t)$.

To turn the LWP $\varphi_{2LW}(\mathbf{r}_1, t)$ into the CLWP $\varphi_{2CL}(\mathbf{r}_1, t)$ two amendments have to be carried out:

a) *Taking the velocity of the observer in the correction term into account.* When the observer (in our case the charge $q_1$) is not at rest, i.e. $\mathbf{v}_1 \neq 0$ then the relative velocity of the interacting charges $\mathbf{v}(t_{r2}) = \mathbf{v}_1(t_{r2}) - \mathbf{v}_2(t_{r2})$ takes the place of $-\mathbf{v}_2(t_r)$ in the correction term as shown in (2.8). This change of velocities is a simply elimination of the arbitrary restriction in LWP.

b) *Correction the length of the retardation time.* The change of the length of the retardation time is a consequence of the motion of the observer, $q_1$. In the case of a moving $q_1$ the electric action performs the path $s_2(t_{r2})$ when propagating from $q_2$ to $q_1$. Here $\mathbf{s}_2(t_{r2}) = \mathbf{r}_1(t) - \mathbf{r}_2(t_{r2})$, see figure 2. Therefore $t - t_{r2} = s_2(t_{r2})/c$ takes the place of $t - t_r = r(t_r)/c$ introduced in (2.3). This way we got the CLW potential $\varphi_{2CL}(\mathbf{r}_1, t)$:

$$\varphi_{2CL}(\mathbf{r}_1, t) = \frac{kq_2}{r(t_{r2}) + \frac{1}{c} \frac{s_2(t_{r2})}{r(t_{r2})} \mathbf{v}(t_{r2}) \mathbf{r}(t_{r2})}, \quad \text{where} \quad t_{r2} = t - \frac{s_2(t_{r2})}{c}. \tag{3.1}$$

At first sight CLWP implies the relative position $\mathbf{r}(t_{r2})$, velocity $\mathbf{v}(t_{r2})$ and distance $r(t_{r2})$ of the interacting charges. Actually, $s_2(t_{r2}) = (t - t_{r2})c$ implies the absolute velocity $\mathbf{v}_2(t_{r2})$ of the source charge, $q_2$ implicitly. Therefore CLWP is not a Galilean invariant potential in this form.

### 3.2 The second order approximation of CLWP

The objective of the second order approximation of CLWP is to get the potential $^2\varphi_{2CL}(\mathbf{r}_1, t)$ in an explicit form and expressed it by the present time values of the measures of $\mathbf{r}(t)$, $\mathbf{v}(t)$, $r(t)$, $s_2(t)$, like an action-at-a distance interaction in appearance. The second order approximation of (3.1);

$$^2\varphi_{2CL}(\mathbf{r}_1, t) = \frac{kq_2}{r} \left[ 1 + \frac{v^2}{2c^2} + \frac{\mathbf{a}\mathbf{r}}{2c^2} - \frac{(\mathbf{r}\mathbf{v})^2}{2c^2 r^2} \right], \tag{3.2}$$

The second order approximation of CLWP cancels the terms of absolute velocity and a Galilean invariant potential is obtained. Unfortunately, the second order force law obtained from this potential does not fit the experience.

### 3.3 The construction of the retarded Coulomb potential (RCP)

The last step to get RCP is a refinement of the correction term in (3.1). The correction term includes the scalar product $\mathbf{v}(t_{r2})\mathbf{r}(t_{r2})$ multiplied by the retardation time of $s_2(t_{r2})/c$. Actually, the scalar product changes during the retardation time, because the interacting charges move and their relative position changes. To get a more accurate result let us take the change of $\mathbf{v}(t_{r2})\mathbf{r}(t_{r2})$ into account and replace the multiplication by $s_2(t_{r2})/c$ by a time integral. Because, $s_2(t_{r2})/c = (t - t_{r2})$;

$$\frac{s_2(t_{r2})}{c} \mathbf{v}(t_{r2}) \mathbf{r}(t_{r2}) \rightarrow \int_{t_{r2}}^{t} \mathbf{v}(t') \mathbf{r}(t') dt' = \frac{r^2(t)}{2} - \frac{r^2(t_{r2})}{2}, \tag{3.3}$$

Replacing (3.3) into (3.1) the RCP is obtained in its final form:

$$\varphi_{2RC}(\mathbf{r_1}, t) = \frac{kq_2}{\frac{r(t_{r2})}{2} + \frac{r^2(t)}{2r(t_{r2})}}. \tag{3.4}$$

The RC potential $\varphi_{2RC}(\mathbf{r_1}, t)$ formulated in (3.4) is the retarded electric potential of the moving charge $q_2$ at the location of another moving charge $q_1$ at the present moment of $t$. Because $t_{r2}$ is defined by $t_{r2} = t - s_2(t_{r2})/c$ and $s_2(t_{r2})$ implies the absolute velocity $\mathbf{v_2}$, the potential (3.4) is not Galilean invariant.

*3.4 The second order, action-at-a-distance approximation of the RC potential and the retarded electric force*

The RC potential $\varphi_{2CR}(\mathbf{r_1}, t)$ depends on the variable, $t_{r2}$ on the RHS of equation (3.4). The retarded distance $r(t_{r2})$, have to be expressed by means of the values at the moment of perception $\mathbf{r}(t)$, $\mathbf{v}(t)$, $r(t)$, $\mathbf{v_2}(t)$. This transformation of variables will be performed by developing the $r(t_{r2})$, and $s_2(t_{r2})$ quantities into power series with the second order term of $1/c$ inclusive. The derivation of the second order approximation of RCP is located in the Appendix.

Omitting the argument $(t)$ in the $\mathbf{r}(t)$, $\mathbf{v}(t)$, $r(t)$ functions the second order approximation of the RCP (3.4);

$$^2\varphi_{1CR}(\mathbf{r_2}, t) = \frac{kq_1}{r}\left[1 - \frac{(\mathbf{rv})^2}{2r^2c^2}\right], \qquad \text{and} \quad ^2\varphi_{2CR}(\mathbf{r_1}, t) = \frac{kq_2}{r}\left[1 - \frac{(\mathbf{rv})^2}{2r^2c^2}\right], \tag{3.5}$$

as obtained in (A.7). The second order approximations of the retarded potentials are seemingly action-at-a-distance potentials because are expressed by the present time measures. The terms of the absolute velocities of $\mathbf{v_1}$ and $\mathbf{v_2}$ are canceled out in the second order approximation. Only the relative quantities remain on the RHS. Therefore, the second order potentials (3.5) are Galilean invariant.

Now, it is possible to write the retarded electric interaction energy of the moving charges, as

$$^2\mathcal{E}(\mathbf{r}, \mathbf{v}) = q_1\,^2\varphi_{2R}[\mathbf{r_1}(t)] = q_2\,^2\varphi_{1R}[\mathbf{r_2}(t)] = \frac{kq_1q_2}{r}\left[1 - \frac{(\mathbf{rv})^2}{2r^2c^2}\right]. \tag{3.6}$$

(3.6) is identical with the Weber's interaction energy, see Assis and Torres [8]. The derivative of the interaction energy with respect to the position vector of the object on which the force is acting provides the force law,

$$^2\mathbf{F_{1RC}} = -\frac{d\,^2\mathcal{E}(\mathbf{r}, \mathbf{v})}{d\mathbf{r_1}}, \qquad\qquad ^2\mathbf{F_{2RC}} = -\frac{d\,^2\mathcal{E}(\mathbf{r}, \mathbf{v})}{d\mathbf{r_2}}. \tag{3.7}$$

The force law obtained is a second order, action-at-a-distance approximation. Because, $\mathbf{r} = \mathbf{r_1} - \mathbf{r_2}$, it is obvious, that $\frac{d}{d\mathbf{r_1}} = \frac{d}{d\mathbf{r}}$, and $\frac{d}{d\mathbf{r_2}} = -\frac{d}{d\mathbf{r}}$. The derivation $\frac{d\,^2\mathcal{E}(\mathbf{r},\mathbf{v})}{d\mathbf{r}}$ is easy to perform as, $\frac{d\,^2\mathcal{E}(\mathbf{r},\mathbf{v})}{d\mathbf{r}} = \frac{dr}{d\mathbf{r}}\frac{dt}{dr}\frac{d\,^2\mathcal{E}(\mathbf{r},\mathbf{v})}{dt}$. The sought, retarded electric force law obtained, as the well known Weber's force law;

$$^2\mathbf{F_{1RC}} = -\frac{d\,^2\mathcal{E}(\mathbf{r}, \mathbf{v})}{d\mathbf{r_1}} = \frac{kq_1q_2}{r^3}\mathbf{r}\left[1 + \frac{v^2}{c^2} - \frac{3(\mathbf{vr})^2}{2r^2c^2} + \frac{\mathbf{ar}}{c^2}\right] \equiv \mathbf{F_{1W}}. \tag{3.8}$$

The same result comes out by means of the Euler-Lagrange equation, when considering $U = kq_1q_2/r$ to be the potential and $T = -kq_1q_2(\mathbf{rv})^2/(2r^3c^2)$ the kinetic part of the (3.6) interaction energy and the Lagrangian is given by $L = T - U$.

The two terms in the square bracket of (3.8) including the relative velocity $v^2$ and $(\mathbf{vr})^2$ represent the magnetic forces. The term of acceleration $\mathbf{r}(\mathbf{ra}) \equiv \mathbf{r} \times (\mathbf{r} \times \mathbf{a}) + \mathbf{a}r^2$ provides the correct radiation term of the accelerated charge, see [14] and $\mathbf{a}r^2$ is the electric dragging term.

## 4    Discussion and conclusion

### 4.1 Weber's force derived from a retarded potential

Weber built his Galilean invariant force law (3.8) upon experimental data and declared the EM force to be an action-at-a-distance force, as it was conventional that time. In this paper Weber's law was derived from the second order approximation of the RCP, hence Weber's force is a retarded force. Although Weber considered his law to be an exact relationship, it is an approximation of $1/c^2$. Present rediscovery gives Weber's force law an important theoretical background.

Through the success of Maxwell's field theory and SR and through its action-at-a-distance features Weber's force law disappeared from the textbooks. This study raises Weber's force law to its rightful place with a delay of 175 years.

### 4.2 The meaning of Weber's force law

The retarded electric force of moving charges obtained from the second order approximation of RCP includes the magnetic forces as well. However, magnetism must be the retarded electricity in itself. There is no need to introduce extra magnetic vector potential to find magnetic forces. Magnetic forces come simply from the retarded Coulomb potential (RCP). A similar standpoint was taken by Mende [15] who doubted the reason for existence of the notion of magnetism in his scalar-vector theory of electricity. Present calculations confirm his doubt.

The force acting between a current carrying conductor and a point charge at rest predicted by Weber is hard to detect because two other effects; electric polarization and electric field inside the conductor, give rise to forces by order of magnitudes stronger [7]. There have not been any experiment carried out which could decide the existence of the force in question.  But the same force was found [15] in a different way.

### 4.3 The third order approximation of RCP and EM waves

Some efforts were made by different authors to extend Weber's potential for radiation of EM waves. Wesley [16] introduced two additional potentials to the usual electric and magnetic potentials to solve the problem of EM radiation.

In present study the second order approximation of RCP lead to a Galilean invariant potential and Weber's force law. Weber's law implies the term of radiation of accelerating charges [14], does not conflict with the existence of EM waves. Maxwell's electrodynamics provides EM waves in the third order approximation of the potential.  The third order approximation of RCP implies terms which depend on the absolute velocity and acceleration of the interacting charges. These Maxwell type third order terms maintain EM waves, as preliminary calculations indicate. The RCP introduced in this paper offers the possibility to develop a complete electrodynamics which implies the Galilean invariant EM interaction of moving charges and the existence of absolute EM waves. The third order approximation has to perform carefully. A model of the propagation of light in Euclidean geometry is needed.

### 4.4  Conclusions

Based on the result of this study the next assertions can be made:

There exists a Galilean invariant formulation of EM interaction of moving charges in a second order approximation based on the finite propagation speed of EM action. This formulation is the Weber's law.

Magnetic forces are the manifestation of the retarded electricity exclusively in Euclidean geometry.

The physical equivalence of Weber's law and a, from SR calculated second order force remain in question.

**Appendix – The second order approximation of the RCP**

To generate the second order approximation of the RCP $\varphi_{2RC}[\mathbf{r}_1, t]$ given in (3.4);

$$\varphi_{2CR}[\mathbf{r}_1, t] = \frac{2kq_2}{r(t_{r2}) + r^2(t)\frac{1}{r(t_{r2})}}, \qquad (A.1)$$

the quantity $r(t_{r2})$ has to be expressed by the measures of $\mathbf{r}(t)$, $\mathbf{v}(t)$, $r(t)$, $\mathbf{a}(t)$, taken at the moment of the perception, $t$. Here $t_{r2} = t - s_2(t_{r2})/c$ as it follows from (3.1). Let us first write the second order approximation of the denominator of (A.1) in the form;

$$^2N_{2R} = {}^2r(t_{r2}) + r^2(t)\ ^2\left[\frac{1}{r(t_{r2})}\right]. \qquad (A.2)$$

Here one needs the second order approximation of the power series of $^2r(t_{r2})$ at the time $t$,

$$^2r(t_{r2}) = r(t) + \dot{r}(t)(t_{r2} - t) + \frac{\ddot{r}(t)(t_{r2} - t)^2}{2}, \quad \text{where} \quad \dot{r} = \frac{\mathbf{r}\mathbf{v}}{r}, \qquad \ddot{r} = \frac{v^2}{r} + \frac{\mathbf{a}\mathbf{r}}{r} - \frac{(\mathbf{r}\mathbf{v})^2}{r^3}. \qquad (A.3)$$

Omitting the argument $(t)$ and introducing the notations $x = t_{r2} - t$, $\alpha = \dot{r}/r$, further, $\beta = \ddot{r}/r$,

$$^2r(t_{r2}) = r\left(1 + \alpha x + \beta\frac{x^2}{2}\right) \quad \text{and} \quad ^2\left[\frac{1}{r(t_{r2})}\right] = \frac{1}{r}\left(1 - \alpha x - \beta\frac{x^2}{2} + \alpha^2 x^2\right) \qquad (A.4)$$

Replacing (A.4) into (A.2) we obtain: $\qquad ^2N_{2R} = r(2 + \alpha^2 x^2) \qquad (A.5)$

The second order approximation of the RCP after replacing the original meaning of $\alpha$ and $x$ and inserting $(t_{r2} - t)^2 = s_2^2(t_{r2})/c^2$ into (A.5) and at last (A.5) into (A.1):

$$^2\varphi_{2CR}[\mathbf{r}_1, t] = \frac{2kq_2}{^2N_{2R}} = \frac{kq_2}{r\left(1 + \frac{\alpha^2 x^2}{2}\right)} = \frac{kq_2}{r\left(1 + \frac{\dot{r}^2 s_2^2(t_{r2})}{2r^2 c^2}\right)}. \qquad (A.6)$$

The second term in the denominator of (A.6) is a second order expression in $1/c$. So, one needs only the zero order approximation of $s_2(t_{r2})$ denoted by $^0s_2(t_{r2})$ to insert into (A.6). It is easy to see from the geometry (figure 2.) that $^0\mathbf{s}_2(t_{r2}) = \mathbf{r}_1 - {}^0\mathbf{r}_2(t_{r2}) = \mathbf{r}$. From this follows; $\left|{}^0\mathbf{s}_2(t_{r2})\right| = {}^0s_2(t_{r2}) = r$.

Inserting the expression $^0s_2(t_{r2}) = r$ into (A.6) the second order approximation of RCP is obtained:

$$^2\varphi_{2CR}[\mathbf{r}_1, t] = \frac{kq_2}{r\left(1 + \frac{\dot{r}^2}{2c^2}\right)} = \frac{kq_2}{r}\left(1 - \frac{\dot{r}^2}{2c^2}\right) = \frac{kq_2}{r}\left[1 - \frac{(\mathbf{r}\mathbf{v})^2}{2r^2 c^2}\right]. \qquad (A.7)$$

**Acknowledgements**

I am grateful prof. Assis for calling my attention to Weber's electrodynamics and for the impressive conversation when we met in Budapest.